\documentclass[12pt]{article}

\usepackage{amsmath}
\usepackage{amssymb}

\begin{document}

\title{Rigidly Supersymmetric Gauge Theories on
Curved Superspace}

\author{Bei Jia, Eric Sharpe\\
        \footnotesize{Department of Physics, Virginia Tech,
Blacksburg, VA 24061 USA}\\
\footnotesize{{\tt beijia@vt.edu}, {\tt ersharpe@vt.edu}}}

\date{}

\maketitle

\begin{abstract}
In this note we construct rigidly supersymmetric gauged sigma
models and gauge theories on certain Einstein four-manifolds,
and discuss constraints on
these theories.  In work elsewhere,
it was recently shown that on some nontrivial Einstein four-manifolds
such as AdS$_4$, $N=1$ rigidly supersymmetric
sigma models are constrained to have
target spaces with exact K\"ahler forms.  Similarly, in gauged
sigma models and gauge theories, we find that supersymmetry
imposes constraints on Fayet-Iliopoulos
parameters, which have the effect of enforcing that K\"ahler forms on
quotient spaces be exact.  We discuss the `background principle' in
this context.  We also discuss general aspects of universality
classes of gauged sigma models, as encoded by stacks, and also discuss
affine bundle structures implicit in these constructions.
In an appendix, we discuss how anomalies in
four-dimensional gauge theories, such as those which play an important
role in our analysis, can be recast in the language of stacks.
\end{abstract}

\begin{flushleft}
September 2011
\end{flushleft}

\newpage

\tableofcontents

\newpage

\section{Introduction}

The idea of supersymmetry has been around for nearly forty years,
which generated numerous discussions ranging from model building in
particle physics to pure theoretical investigations,
and proved to be a powerful tool for understanding quantum field theory.
Historically, most discussions of four-dimensional rigidly supersymmetric
nonlinear sigma models have focused on Minkowski spacetimes.
Recently, rigidly supersymmetric nonlinear sigma models on some
nontrivial four-dimensional spacetime manifolds have been discussed by
several groups \cite{allanetal,fest-seib,butter-kuzenko}.
(See also references contained therein for older literature on this
subject.)
These have
interesting new properties,
different from the traditional Minkowski spacetime models, essentially
because one must add additional terms to the action to take into account
the curvature of the spacetime manifold.

One way to derive those extra terms in the action is to manually
add extra terms consistent with the requirements imposed by
(rigid) supersymmetry
\cite{allanetal,butter-kuzenko}.
Another approach \cite{allanetal,fest-seib} is to
start with a supergravity theory in four
dimensions, then decouple gravity in order to obtain a theory that
is rigidly supersymmetric.
Demanding that the supersymmetry variation of the gravitino vanishes
then constrains the possible spacetime four-manifolds.
The solutions of these constraining equations generate two classes
of spacetime geometries, including AdS$_4$ and $S^4$ (after Wick rotation
to a Euclidean metric) in one class, and $S^3 \times \mathbb{R}$ in a second
class that requires a covariantly constant vector field.

These theories have many interesting properties that are different from
the Minkowski spacetime case.  For instance \cite{allanetal,fest-seib},
the target spaces of the
supersymmetric nonlinear sigma models on spacetimes such as AdS$_4$ and $S^4$
must be noncompact K\"{a}hler manifolds, with exact K\"ahler forms.
Furthermore,
the Lagrangian depends
only on certain combination of the K\"{a}hler potential and the superpotential
 -- neither alone is physically meaningful.

In this paper we construct
$N=1$ rigidly
supersymmetric gauged nonlinear sigma models and gauge theories on nontrivial
four-dimensional spacetime manifolds, by starting with $N=1$ supergravity
and decoupling gravity.
Just as target spaces of rigidly supersymmetric ungauged
theories are constrained, we find analogous constraints in
gauge theories.  For example, just as the Fayet-Iliopoulos
parameter is constrained in $N=1$ supergravity
\cite{dist-sharpe,hellerman-sharpe,seib-may},
we find
a constraint on the Fayet-Iliopoulos parameter in rigidly supersymmetric
theories, which guarantees that
the K\"ahler form on the quotient space is exact.
Just as in $N=1$ supergravity in four dimensions, the superpotential is
a section of a line bundle \cite{bagger-witten}, we interpret the
superpotential in these rigidly supersymmetric theories as a section
of an affine bundle.
Just as in
$N=1$ supergravity \cite{dist-sharpe,hellerman-sharpe},
where the Fayet-Iliopoulos parameter was determined
by the group action on the Bagger-Witten line bundle,
here too the Fayet-Iliopoulos parameter can be understood
in terms of lifts to the affine bundle.

We should mention that some analogous results were obtained in linearized
supergravity theories obtained by coupling a rigidly supersymmetric
theory to gravity.  In such theories, for {\it e.g.}
couplings involving the Ferrara-Zumino multiplet,
one also often sees that K\"ahler forms are exact and
Fayet-Iliopoulos parameters vanish \cite{zohar1,zohar2},
just as we
describe here for rigidly supersymmetric theories on {\it e.g.}
AdS$_4$.  (As observed in \cite{seib-may}, however, one should distinguish
supergravities obtained by coupling a rigid theory to gravity, from
more general supergravity theories.  For example, in generic heterotic Calabi-Yau compactifications to four
dimensions and N=1 supergravity,
it is widely believed that the
Bagger-Witten line bundle is nontrivial, and so such supergravities
cannot be obtained by coupling a rigid theory in the fashion above.)

We start in
section~\ref{sect:review}
with a review of those rigidly
supersymmetric nonlinear sigma models constructed in
\cite{allanetal,fest-seib,butter-kuzenko}, using the superspace formulation
to make the story more compact. We also give the interpretation of the
superpotential as a section of certain affine bundle over the target space.
In section~\ref{sect:gauge} we construct supersymmetric gauged sigma models
and gauge
theories, from which we derive some constraints on the theory.
We find that the Fayet-Iliopoulos parameter has to vanish in these theories,
which has the effect of enforcing that K\"ahler forms on quotient spaces
be exact. In section~\ref{sect:stacks} we discuss some general aspects of
universality classes
of gauged sigma models, as encoded by stacks, and theories defined by
restrictions on nonperturbative sectors.
We also provide some mathematical background
about affine bundles and equivariant structures on affine bundles in
an appendix.

In passing, since one of the spaces we study will be AdS$_4$,
we should mention that, just as in previous papers
\cite{allanetal,fest-seib}, we shall
ignore the role of boundary conditions.
See {\it e.g.} \cite{mv,iw,ms,amr}
for an overview of boundary
conditions in AdS$_4$, and {\it e.g.} \cite{cw,gkrrs,rr1} for information on
how such boundary conditions can restrict chiral matter representations.

\section{Review of rigidly susy sigma models on curved superspace}
\label{sect:review}

There are several ways of deriving rigid supersymmetric nonlinear sigma model
from supergravtiy. For example, we can decouple gravity in the weak coupling
limit to get supersymmetric nonlinear sigma model on AdS$_4$ \cite{allanetal}.
On the other hand, it was noted in \cite{fest-seib} that the auxiliary fields
$b_{\mu}$
and $M$ from the $N=1$ supergravity multiplet could be used to determine
the geometry of spacetime, therefore generating two classes of spacetime
geometries. The idea is to start with $N=1$ supergravity Lagrangian, then
set the gravitino to zero to completely remove the dynamics of gravity,
and make the auxiliary fields $b_{\mu}$ and $M$ from the supergravity multiplet
satisfy certain constraining equations to make sure we have
$N=1$ supersymmetry, as well as the ability to perform a modified
K\"{a}hler transformation with the resulting Lagrangian invariant.

Let us review the approach of \cite{fest-seib}, as we shall apply it to
gauge theories in the next section.
We start with the $N=1$ chiral supergravity Lagrangian in
superspace \cite{wb}:
\begin{equation}
\mathcal{L}=\frac{1}{\kappa^2}\int d^2\Theta \, 2\mathcal{E}
\left[\frac{3}{8}(\bar{\mathcal{D}}\bar{\mathcal{D}}-8R)
\exp(-\frac{\kappa^2}{3}K(\Phi^i, \bar{\Phi}^{\bar{\imath}}))+\kappa^2 W(\Phi^i)
\right]+h.c.
\end{equation}
Then we remove the effect of gravity.
First, we need to expand in $\kappa^2$, then only keep the terms that are
independent of $\kappa$. We get
\begin{equation}
\mathcal{L}=\int d^2\Theta \, 2\mathcal{E}
\left[-\frac{1}{8}(\bar{\mathcal{D}}\bar{\mathcal{D}}-8R)K(\Phi^i,
\bar{\Phi}^{\bar{\imath}})+ W(\Phi^i)\right]+h.c.
\end{equation}

As observed in \cite{fest-seib}, in order for consistency of the method above,
we must demand that the spacetime be such that the supersymmetry variation
of the gravitino vanishes, as we have truncated it.
The off-shell supersymmetry variation of the gravitino is of the form
\cite{fest-seib, wb}
\begin{equation}
\begin{split}
& \delta \Psi_{\mu}^{\alpha}  =
-2 \nabla_{\mu} \zeta^{\alpha} + \frac{i}{3}
\left( M (\epsilon\sigma_{\mu} \overline{\zeta} )^{\alpha} +
3 b_{\mu} \zeta^{\alpha}  +  2b^{\nu} ( \zeta \sigma_{\nu \mu} )^{\alpha}
\right),\\
& \delta \overline{\Psi}_{\mu\dot{\alpha}}  =
-2 \nabla_{\mu} \overline{\zeta}_{\dot{\alpha}}  -  \frac{i}{3}
\left( \overline{M} (\zeta \sigma_{\mu} )_{\dot{\alpha}} +
3 b_{\mu} \overline{\zeta}_{\dot{\alpha}}  +
2b^{\nu} ( \overline{\zeta} \overline{\sigma}_{\nu \mu} )_{\dot{\alpha}}
\right),
\end{split}
\end{equation}
and demanding that it vanishes implies constraints
of the form \cite{fest-seib}[equ'n (2.11)]
\begin{displaymath}
\begin{array}{c}
M b_{\mu} \: = \: \bar{M} b_{\mu} \: = \: 0, \: \: \:
\nabla_{\mu} b_{\nu} \: = \: 0, \: \: \:
\partial_{\mu} M \: = \: \partial_{\mu} \bar{M} \: = \: 0, \: \: \:
W_{\mu \nu \kappa \lambda} \: = \: 0, \\
R_{\mu \nu} \: = \:
- \frac{2}{9}\left( b_{\mu} b_{\nu} \: - \: g_{\mu \nu} b_{\rho} b^{\rho}
\right) \: + \: \frac{1}{3} g_{\mu \nu} M \bar{M},
\end{array}
\end{displaymath}
where $M$, $\bar{M}$, $b_{\mu}$ are auxiliary fields in the
N=1 supergravity multiplet, and $W_{\mu \nu \kappa \lambda}$
is the Weyl tensor.
According to \cite{fest-seib},
there are two classes of solutions to the equations above, namely:
\begin{enumerate}
\item $b_{\mu} = 0$, constant $M$, $\bar{M}$,
\item $M = \bar{M} = 0$, $b_{\mu}$ a covariantly-constant vector.
\end{enumerate}
In the first case, if we Wick rotate to Euclidean space, it can be
argued from the existence of spinors $\zeta$ in the gravitino variation
\cite{bryant-priv} that the spacetime metric either has constant sectional
curvature, or is Ricci-flat and self-dual or anti-self-dual. To see this, note
in this case we have Killing spinor equations
\begin{equation}
\begin{split}
& \nabla_{\mu} \zeta^{\alpha} = \frac{i}{6} M (\epsilon\sigma_{\mu} \bar{\zeta} )^{\alpha},\\
& \nabla_{\mu} \bar{\zeta}_{\dot{\alpha}}  =  \frac{-i}{6}\bar{M} (\zeta \sigma_{\mu} )_{\dot{\alpha}},
\end{split}
\end{equation}
Taking the covariant derivative of the first equation, then together with the second equation we can get
\begin{equation}
\begin{split}
\nabla_{\nu}\nabla_{\mu} \zeta^{\alpha} = \frac{1}{36} M \bar{M}(\sigma_{\mu}\zeta\bar{\sigma}_{\nu})^{\alpha},
\end{split}
\end{equation}
then it follows that
\begin{equation}
\begin{split}
[\nabla_{\mu},\nabla_{\nu}] \zeta^{\alpha} = R_{\mu\nu\rho\sigma}(\zeta\sigma^{\rho\sigma})^{\alpha}
= \frac{1}{9} M \bar{M}(\zeta\sigma_{\mu\nu})^{\alpha},
\end{split}
\end{equation}
Similarly, we can find an analogous equation from the second Killing spinor equation above
\begin{equation}
\begin{split}
[\nabla_{\mu},\nabla_{\nu}] \bar{\zeta}_{\dot{\alpha}} = R_{\mu\nu\rho\sigma}(\bar{\zeta}\bar{\sigma}^{\rho\sigma})_{\dot{\alpha}}
= \frac{1}{9} M \bar{M}(\bar{\zeta}\bar{\sigma}_{\mu\nu})_{\dot{\alpha}},
\end{split}
\end{equation}
Suppose we have $\zeta^{\alpha}\neq 0$ as well as $\bar{\zeta}_{\dot{\alpha}}\neq 0$, then we see
\begin{equation}
R_{\mu\nu\rho\sigma}\sigma^{\rho\sigma}\propto g_{\mu\rho}g_{\nu\sigma}\sigma^{\rho\sigma},
\end{equation}
which is equivalent to
\begin{equation}
R_{\mu\nu\rho\sigma}\sigma^{\rho\sigma}\propto g_{\mu\sigma}g_{\nu\rho}\sigma^{\sigma\rho}=-g_{\mu\sigma}g_{\nu\rho}\sigma^{\rho\sigma},
\end{equation}
Therefore, from linear independence of the $\sigma$'s, we see that
\begin{equation}
R_{\mu\nu\rho\sigma}\propto g_{\mu\rho}g_{\upsilon\sigma}
- g_{\mu\sigma}g_{\nu\rho},
\end{equation}
which means the spacetime has constant sectional curvature, {\it i.e.}
it is a space form.
Similarly, if $\zeta^{\alpha}=0$ (or $\bar{\zeta}_{\dot{\alpha}}=0$),
it follows that the spacetime is Ricci-flat and self-dual (or anti-self-dual).
In the second case, it can be similarly argued \cite{bryant-priv}
that the spacetime metric is a product of a line and a metric of
nonnegative constant sectional curvature.

\subsection{$M = \bar{M} = $constant, $b_{\mu}=0$}

Now, let us specialize to the first case, in which
$M$, $\bar{M}$ are nonzero constants, and $b_{\mu}$ vanishes.
With this choice the above superspace Lagrangian can be written
in an interesting form
\begin{equation}   \label{ungauged-sigma-model-superspace}
\mathcal{L}=\int d^2\Theta \, 2\mathcal{E}
\Big[-\frac{1}{8}(\bar{\mathcal{D}}\bar{\mathcal{D}}-8R)
\big(K(\Phi^i, \bar{\Phi}^{\bar{\imath}})-
\frac{3}{2M}W(\Phi^i)-\frac{3}{2\bar{M}}\bar{W}(\bar{\Phi}^{\bar{\imath}})
\big)\Big]+h.c.
\end{equation}
from which we can clearly see that this nonlinear sigma model Lagrangian
depends explicitly on the combination of the K\"{a}hler potential and the
superpotential
\begin{equation}  \label{kwcombo}
K \: - \: \frac{3}{2M}W \: - \: \frac{3}{2\bar{M}}\bar{W},
\end{equation}
which suggests that $K$ and $W$ alone
are not physically meaningful; it is only the combination above
that is physically meaningful.
This makes the modified K\"{a}hler transformation mentioned
above apparent, which
can be derived as following: if we perform super-K\"{a}hler transformation
to the superspace K\"{a}hler potential
\begin{equation}
K(\Phi,\bar{\Phi})\mapsto K(\Phi,\bar{\Phi})+F(\Phi)+\bar{F}(\bar{\Phi})
\end{equation}
then in order to make the superspace
Lagrangian~(\ref{ungauged-sigma-model-superspace}) invariant,
we must transform the superspace superpotential $W(\Phi)$ accordingly,
which leads the the following transformation of the superpotential
\begin{equation}
W(\Phi) \mapsto W(\Phi) +\frac{2}{3}M F(\Phi).
\end{equation}
The Lagrangian is invariant under the combination of these two transformation
(the modified K\"{a}hler transformation)
\footnote{Note in supergravity, the superspace Lagrangian is invariant under
the combined super-K\"{a}hler and super-Weyl transformations,
the latter of which is a transformation of the superspace superpotential
\cite{wb} which indicates the fact that the superpotential is a holomorphic
section of a line bundle over the target space \cite{bagger-witten,wb}}.
This  superspace transformation leads exactly to
what was observed by others \cite{allanetal, fest-seib,butter-kuzenko} that the nonlinear sigma model action is invariant under the K\"{a}hler
transformation of the target space $X$
\begin{equation}  \label{kahler-k-trans}
K(\phi,\bar{\phi})\mapsto K(\phi,\bar{\phi})+f(\phi)+\bar{f}(\bar{\phi})
\end{equation}
supplemented by the following transformation of the superpotential
\begin{equation}  \label{kahler-w-trans}
W(\phi)\mapsto W(\phi)+\frac{2}{3}M f(\phi).
\end{equation}
Let $U_{\alpha}$ and $U_{\beta}$ be two open subsets of the target space $X$.
Then across $U_{\alpha}\cap U_{\beta}$ we have
\begin{equation}  \label{kahler-cech}
\begin{split}
K_{\alpha}&\mapsto K_{\beta}+f_{\alpha\beta}+\bar{f}_{\alpha\beta},\\
W_{\alpha}&\mapsto W_{\beta}+\frac{2}{3}M f_{\alpha\beta},
\end{split}
\end{equation}
which is a clear indication that the superpotential $W$ is not a function
globally on $X$, but rather is a section of a rank 1 affine bundle\footnote{
See appendix~\ref{app:affine} for a discussion of affine bundles.
} $(\mathcal{O},A)$ over $X$,
whose line bundle part is trivially $\mathcal{O}$,
while the ${\cal O}$-torsor $A$
is determined by the geometry of the spacetime and the
K\"{a}hler transformation of the target space.
Then the combination~(\ref{kwcombo}) should be interpreted as a pairing between sections of affine bundles and their dual bundles,
which is globally well-defined and invariant under
K\"{a}hler transformation of the target space.

Physically, the transformations above mean that there is not a well-defined
global function $W$ that we can think of as the superpotential,
as suggested by the appearance of the combination~(\ref{kwcombo}).
We can combine $(K_{\alpha}, W_{\alpha})$ on patches into
\begin{displaymath}
\left(K'_{\alpha} \equiv K_{\alpha} -\frac{3}{2M}W_{\alpha}-\frac{3}{2\bar{M}}
\bar{W}_{\alpha}, 0\right),
\end{displaymath}
and then
perform another K\"ahler transformation to
\begin{displaymath}
\left(K'_{\alpha} -\frac{3}{2M} f_{\alpha}-\frac{3}{2\bar{M}} \bar{f}_{\alpha},
f_{\alpha}\right),
\end{displaymath}
thus replacing $W_{\alpha}$ by $f_{\alpha}$,
for any holomorphic function $f_{\alpha}$ on the patch.  Only the combination
of $K$ and $W$ is physically meaningful.

One consequence of this phenomenon is that the target space $X$ is
necessarily noncompact ~\cite{allanetal, fest-seib,butter-kuzenko}.
Equation~(\ref{kahler-cech}) not only requires
the \u{C}ech cocycle
$(\delta f)_{\alpha\beta\gamma}=0$ on all triple overlaps,
but also requires that the \u{C}ech cocycle be trivial ($f_{\alpha \beta}$
is a \u{C}ech coboundary),
therefore the K\"{a}hler form of $X$ must be cohomologically trivial,
which leads to the noncompactness of $X$,
as well as the existence of a globally defined K\"{a}hler potential.\\
\\
\textbf{Example 1: AdS$_4$}\\
To describe the spacetime AdS$_4$ spacetime, we set
\begin{equation} \label{AdS}
\begin{split}
M&=\bar{M}=- \frac{3}{2r},\\
b_{\mu}&=0,
\end{split}
\end{equation}
where $r$ can be interpreted as the radius of the AdS$_4$ curvature,
with the scalar curvature given by $\mathcal{R}=\frac{15}{2r^2}$.
\footnote{Two notes on conventions.
First, we are working in mostly-plus metric conventions in this paper,
same as \cite{wb},
and in those conventions, typically the AdS$_4$ curvature is negative,
not positive.  The reason it is positive above is that we are following
the conventions of \cite{wb}, in which the spin connection has an
atypical sign \cite{wb}[equ'n (17.12)],
and which results in the AdS$_4$ curvature
being positive instead of negative.  We would like to thank the authors
of \cite{allanetal} for explaining this to us.
Second, the curvature is related to other descriptions as follows.
If we describe AdS$_4$ as the hypersurface
$-u^2-v^2+x^2+y^2+z^2=-\alpha^2$
in $\mathbb{R}^{2,3}$, then its curvature is
$\mathcal{R}=-\frac{12}{\alpha^2}$. }
The resulting superspace Lagrangian is
\begin{equation}   \label{ungauged-sigma-model-superspace-ads}
\mathcal{L}=\int d^2\Theta \, 2\mathcal{E} \Big[-\frac{1}{8}(\bar{\mathcal{D}}\bar{\mathcal{D}}-8R)
\big(K(\Phi^i, \bar{\Phi}^{\bar{i}})+ rW(\Phi^i)+r\bar{W}(\bar{\Phi}^{\bar{i}})\big)\Big]+h.c.
\end{equation}
When expanded in components, we get the off-shell Lagrangian on AdS$_4$
described in \cite{allanetal, fest-seib}, namely
\begin{equation}  \label{ungauged-sigma-model}
\begin{split}
\mathcal{L}=& -g_{i\bar{\jmath}}\partial_{\mu}\phi^i\partial^{\mu}
\bar{\phi}^{\bar{\jmath}}-ig_{i\bar{\jmath}}\bar{\chi}^{\bar{\jmath}}
\bar{\sigma}^{\mu}\mathcal{D}_{\mu}\chi^i+g_{i\bar{\jmath}}F^i
\bar{F}^{\bar{\jmath}}-F^i(\frac{1}{2}g_{i\bar{\jmath},\bar{k}}
\bar{\chi}^{\bar{\jmath}}\bar{\chi}^{\bar{k}}-\frac{1}{r}(K_i+rW_i))\\
& -\bar{F}^{\bar{\imath}}(\frac{1}{2}g_{j\bar{\imath},k}\chi^j\chi^k-
\frac{1}{r}(K_{\bar{\imath}}+r\bar{W}_{\bar{\imath}}))-
\frac{1}{2r}(K_{ij}+rW_{ij})\chi^i\chi^j
-\frac{1}{2r}(K_{\bar{\imath}\bar{\jmath}}+r\bar{W}_{\bar{\imath}\bar{\jmath}})
\bar{\chi}^{\bar{\imath}}\bar{\chi}^{\bar{\jmath}}\\
 & +\frac{1}{4}g_{i\bar{\jmath},k\bar{l}}\chi^i\chi^k\bar{\chi}^{\bar{\jmath}}
\bar{\chi}^{\bar{l}}+\frac{3}{r^2} (K+rW+r\bar{W}),
\end{split}
\end{equation}
where $\mathcal{D}_{\mu}\chi^i=\partial_{\mu}\chi^i-\omega_{\mu}
\chi^i+\Gamma^i_{jk}
\mathcal{D}_{\mu}\phi^j\chi^k$, with $\omega_{\mu}$ the spin connection
on the AdS$_4$ spacetime.
Note in this way we can also recover the supersymmetry transfromation
of the chiral multiplet on AdS$_4$, simply by setting the gravitino to zero
in the supergravity transformation of the chiral multiplet from chiral supergravity, which leads to
\begin{equation}
\begin{split}
\delta_{\zeta}\phi^i&=\sqrt{2}\zeta\chi^i,\\
\delta_{\zeta}\chi^i&=\sqrt{2}F^i\zeta+i\sqrt{2}\sigma^{\mu}
\bar{\zeta}\partial_{\mu}\phi^i,\\
\delta_{\zeta}F^i&=-\frac{\sqrt{2}}{2r}\zeta\chi^i+i\sqrt{2}\bar{\zeta}
\bar{\sigma}^{\mu}\mathcal{D}_{\mu}\chi^i,
\end{split}
\end{equation}
where the supersymmetry parameter $\zeta$ should satisfy the
Killing spinor equations
\begin{equation}
\begin{split}
(\nabla_{\mu}\zeta)^{\alpha}&
=\frac{i}{2r}(\bar{\zeta}\bar{\sigma}_{\mu})^{\alpha},\\
(\nabla_{\mu}\bar{\zeta})_{\dot{\alpha}}&=\frac{i}{2r}
(\zeta\sigma_{\mu})_{
\dot{\alpha}}.
\end{split}
\end{equation}\\
\\
\textbf{Example 2: $S^4$}\\
Next, let us Wick rotate to a Euclidean spacetime.
Consider the case of $S^4$, where
\begin{equation} \label{S4}
\begin{split}
M&=\bar{M}=- \frac{3i}{2r},\\
b_{\mu}&=0.
\end{split}
\end{equation}
The resulting Euclidean Lagrangian in components is
\begin{equation}  \label{ungauged-sigma-model2}
\begin{split}
\mathcal{L}=& g_{i\bar{\jmath}}\partial_{\mu}\phi^i\partial^{\mu}\bar{\phi}^{
\bar{\jmath}}+ig_{i\bar{\jmath}}\bar{\chi}^{\bar{\jmath}}\bar{\sigma}^{\mu}
\mathcal{D}_{\mu}\chi^i-g_{i\bar{\jmath}}F^i
\bar{F}^{\bar{\jmath}}+F^i(\frac{1}{2}g_{i\bar{\jmath},\bar{k}}
\bar{\chi}^{\bar{\jmath}}\bar{\chi}^{\bar{k}}-\frac{i}{r}(K_i-irW_i))\\
& +\bar{F}^{\bar{\imath}}(\frac{1}{2}g_{j\bar{\imath},k}\chi^j\chi^k-
\frac{i}{r}(K_{\bar{\imath}}-ir\bar{W}_{\bar{\imath}}))+
\frac{i}{2r}(K_{ij}-irW_{ij})\chi^i\chi^j
+\frac{i}{2r}(K_{\bar{\imath}\bar{\jmath}}-ir\bar{W}_{\bar{\imath}\bar{\jmath}})
\bar{\chi}^{\bar{\imath}}\bar{\chi}^{\bar{\jmath}}\\
 & -\frac{1}{4}g_{i\bar{\jmath},k\bar{l}}\chi^i\chi^k\bar{\chi}^{\bar{\jmath}}
\bar{\chi}^{\bar{l}}+\frac{3}{r^2} (K-irW-ir\bar{W}).
\end{split}
\end{equation}
(Note in this case the pertinent combination of $K$, $W$,
is $K-irW-ir\bar{W}$, and because the two terms with $W$ are not
complex conjugates, one could debate whether the symmetry mixing
$K$ and $W$ should properly be termed a K\"ahler transformation.)
As discussed in \cite{fest-seib}, this action is not real.

In an analogous fashion as of the AdS$_4$ case, we can find the supersymmetry
transformations of this Euclidean theory on $S^4$
\begin{equation}
\begin{split}
\delta_{\zeta}\phi^i&=\sqrt{2}\zeta\chi^i,\\
\delta_{\zeta}\chi^i&=\sqrt{2}F^i\zeta+i\sqrt{2}\sigma^{\mu}
\bar{\zeta}\partial_{\mu}\phi^i,\\
\delta_{\zeta}F^i&=-\frac{\sqrt{2}i}{2r}\zeta\chi^i+i\sqrt{2}
\bar{\zeta}\bar{\sigma}^{\mu}\mathcal{D}_{\mu}\chi^i,
\end{split}
\end{equation}
Now the Killing spinor equations become
\begin{equation}
\begin{split}
(\nabla_{\mu}\zeta)^{\alpha}&=-\frac{1}{2r}(\bar{\zeta}
\bar{\sigma}_{\mu})^{\alpha},\\
(\nabla_{\mu}\bar{\zeta})_{\dot{\alpha}}&=-\frac{1}{2r}
(\zeta\sigma_{\mu})_{\dot{\alpha}}.
\end{split}
\end{equation}

\subsection{$M = \bar{M} = 0$, $b_{\mu} \neq 0$}

So far we have only reviewed spacetimes corresponding to nonzero $M$ and
vanishing $b_{\mu}$.
The second class of solutions of the auxiliary fields $M$ and $b_{\mu}$ found in
\cite{fest-seib}, corresponding to a different class of spacetime geometries,
are given by $M=\bar{M}=0$ with $b_{\mu}$ a covariantly constant vector.
For example, the spacetime $S^3 \times \mathbb{R}$ is consistent
with the choices
\begin{equation} \label{S3*R}
\begin{split}
M&=\bar{M}=b_i=0,\\
b_0&=- \frac{3}{r}.
\end{split}
\end{equation}
The corresponding component Lagrangian is
\begin{equation}  \label{nlsm-R*S^3}
\begin{split}
\mathcal{L}=& -g_{i\bar{\jmath}}\partial_{\mu}\phi^i\partial^{\mu}\bar{\phi}^{
\bar{\jmath}}-ig_{i\bar{\jmath}}\bar{\chi}^{\bar{\jmath}}\bar{\sigma}^{\mu}
\mathcal{D}_{\mu}\chi^i+g_{i\bar{\jmath}}F^i
\bar{F}^{\bar{\jmath}}-F^i(\frac{1}{2}g_{i\bar{\jmath},\bar{k}}
\bar{\chi}^{\bar{\jmath}}\bar{\chi}^{\bar{k}}-W_i)\\
 & -\bar{F}^{\bar{\imath}}(\frac{1}{2}g_{j\bar{\imath},k}\chi^j\chi^k-
\bar{W}_{\bar{\imath}})-\frac{1}{2}W_{ij}\chi^i\chi^j
-\frac{1}{2}\bar{W}_{\bar{\imath}\bar{\jmath}}\bar{\chi}^{\bar{\imath}}
\bar{\chi}^{\bar{\jmath}}+\frac{1}{4}g_{i\bar{\jmath},k\bar{l}}\chi^i\chi^k
\bar{\chi}^{\bar{\jmath}}\bar{\chi}^{\bar{l}}\\
 &+\frac{i}{r}(K_i\partial_0\phi^i-K_{\bar{\imath}}\partial_0\bar{\phi}^{
\bar{\imath}})+\frac{1}{2r}g_{i\bar{\jmath}}\chi^i\sigma_0\bar{\chi}^{
\bar{\jmath}},
\end{split}
\end{equation}
where the last line contains some new terms that are different from the
familiar Minkowski spacetime model.
Note these extra terms vanishes at the limit $r\rightarrow\infty$,
so this theory reduces to the Minkowski case as expected.
(See for example \cite{dsen1,dsen2} for further discussions of rigidly
supersymmetric theories on this spacetime.)

In this model the supersymmetry transformations are
\begin{equation}
\begin{split}
\delta_{\zeta}\phi^i&=\sqrt{2}\zeta\chi^i,\\
\delta_{\zeta}\chi^i&=\sqrt{2}F^i\zeta+i\sqrt{2}
\sigma^{\mu}\bar{\zeta}\partial_{\mu}\phi^i,\\
\delta_{\zeta}F^i&=\sqrt{2}\bar{\zeta}^{\dot{\alpha}}(i\mathcal{D}_{\alpha\dot{\alpha}}\chi^{\alpha}-\frac{1}{6}b_{\alpha\dot{\alpha}}\chi^{\alpha}),
\end{split}
\end{equation}
where the supersymmetry parameter $\zeta$ must
satisfy
\cite{fest-seib}
\begin{equation}
\begin{split}
&(\nabla_0\zeta)_{\alpha}+\frac{i}{r}\zeta_{\alpha}=0,\\
&2(\nabla_{i}\zeta)_{\alpha}-\frac{i}{r}(\sigma_i\bar{\sigma}_0
\zeta)_{\alpha}=0.
\end{split}
\end{equation}
In these cases there is no shift symmetry combining the K\"{a}hler potential and superpotential into a single quantity;
the resulting Lagrangian is already K\"{a}hler invariant.
Consequently, many of the conventional powerful methods from theories
on Minkowski spacetime, such as holomorphy arguments, can be applied here.

\section{Rigidly susy gauge theory on curved superspace}
\label{sect:gauge}

Now we apply the method of the last section to the $N=1$ gauged supergravity
Lagrangian in superspace \cite{wb}
\begin{equation}
\begin{split}
\mathcal{L}=\frac{1}{\kappa^2}&\int d^2\Theta \, 2\mathcal{E}
\Big[\frac{3}{8}(\bar{\mathcal{D}}\bar{\mathcal{D}}-8R)
\exp\big(-\frac{\kappa^2}{3}[K(\Phi^i, \bar{\Phi}^{\bar{\imath}})+
\Gamma(\Phi^i, \bar{\Phi}^{\bar{\imath}},V)]\big)\\
&+\frac{\kappa^2}{16g^2}W^{(a)}W^{(a)}+\kappa^2 W(\Phi^i)\Big]+h.c.,
\end{split}
\end{equation}
where in Wess-Zumino gauge
\begin{equation}  \label{gauge-defns}
\begin{split}
\Gamma&=V^{(a)}D^{(a)}+\frac{1}{2}g_{i\bar{\jmath}}X^{i(a)}X^{\bar{\jmath}(b)}
V^{(a)}V^{(b)},\\
W_{\alpha}&=-\frac{1}{4}(\bar{\mathcal{D}}\bar{\mathcal{D}}-8R)(\mathcal{D}_{
\alpha}V-\frac{1}{2}[V,\mathcal{D}_{\alpha}V]).
\end{split}
\end{equation}
$X^{(a)}$ are the holomorphic Killing vectors on the target space $X$ extended
to superfields.
We now study the two different classes of spacetime geometries separately.

\subsection{$M=\bar{M}={\rm constant}$, $b_{\mu}=0$}
\label{sect:rigidsusygauge:ads}

Removing the dynamics of gravity and setting the background fields to
$M=\bar{M}={\rm constant}$, $b_{\mu}=0$ to generate the first class of
spacetime geometries, we find in superspace
\begin{equation}  \label{superspace-gauge-lagrangian}
\begin{split}
\mathcal{L}=&\int d^2\Theta \, 2\mathcal{E}
\Big[-\frac{1}{8}(\bar{\mathcal{D}}\bar{\mathcal{D}}-8R)(K(\Phi^i,
\bar{\Phi}^{\bar{\imath}})+\Gamma(\Phi^i, \bar{\Phi}^{\bar{\imath}},V))+
\frac{1}{16g^2}W^{(a)}W^{(a)}+W(\Phi^i)\Big]+h.c.\\
&
\begin{split}
=\int d^2\Theta \, 2\mathcal{E}
\Big[-\frac{1}{8}(\bar{\mathcal{D}}\bar{\mathcal{D}}-8R)
\big[\big(K(\Phi^i, \bar{\Phi}^{\bar{\imath}})
-\frac{3}{2M}W(\Phi^i)-& \frac{3}{2\bar{M}}\bar{W}(\bar{\Phi}^{\bar{\imath}})
\big)+
\Gamma(\Phi^i, \bar{\Phi}^{\bar{\imath}},V)\big]\\
&+\frac{1}{16g^2}W^{(a)}W^{(a)}\Big]+h.c.
\end{split}
\end{split}
\end{equation}

To obtain a gauge invariant Lagrangian from~(\ref{superspace-gauge-lagrangian}),
we need to impose some constraints. Since in both superspace and in components
the Lagrangian only depends on the combination of the K\"ahler potential and
the superpotential
in~(\ref{kwcombo}), namely,
\begin{displaymath}
K \: - \: \frac{3}{2M} W \: - \: \frac{3}{2 \bar{M}} \bar{W},
\end{displaymath}
it is natural to start with the gauge transformation of this globally
well-defined combination. We apply the superspace gauge transformations
\cite{wb}
\begin{equation} \label{super-gauge-trans}
\begin{split}
\delta \bigg(K-\frac{3}{2M}W-\frac{3}{2\bar{M}}\bar{W}\bigg)&
=\Lambda^{(a)}F^{(a)}+
\bar{\Lambda}^{(a)}\bar{F}^{(a)}-i[\Lambda^{(a)}-\bar{\Lambda}^{(a)}]D^{(a)},\\
\delta \Gamma &=i[\Lambda^{(a)}-\bar{\Lambda}^{(a)}]D^{(a)},
\end{split}
\end{equation}
where $\Lambda^{(a)}$ are the
gauge transformation parameters extended to superfields, and
\begin{equation}
F^{(a)}=X^{(a)}\bigg(K-\frac{3}{2M}W-\frac{3}{2\bar{M}}\bar{W}\bigg)+iD^{(a)}
\end{equation}
is a holomorphic function of the superfields $\Phi^i$.
Applying these gauge transformations to the superspace Lagrangian, we get
\begin{equation}
\delta\mathcal{L}=\int d^2\Theta \, 2\mathcal{E}
\Big[-\frac{1}{8}(\bar{\mathcal{D}}\bar{\mathcal{D}}-8R)(\Lambda^{(a)}F^{(a)}+\bar{\Lambda}^{(a)}\bar{F}^{(a)})\Big]+h.c.
\end{equation}
We demand
that the superspace Lagrangian~(\ref{superspace-gauge-lagrangian})
be invariant under these
gauge transformations.  Using the fact that when the
gravitational fields are decoupled,
the vielbein superfield $2 \mathcal{E}$
has the form
\begin{displaymath}
2\mathcal{E} = (1-\Theta\Theta\bar{M})e,
\end{displaymath}
where $e$ is the veilbein, we are led to the following constraining equations by setting the lowest component, the $\Theta$ component and the $\Theta\Theta$ component of the superspace $\delta\mathcal{L}$ to zero
\begin{equation} \label{super-gauge-trans-2}
\begin{split}
M^2\epsilon^{(a)}F^{(a)}(\phi)& =0\\
MF^i\epsilon^{(a)}\partial_iF^{(a)}(\phi)& =0\\
M\chi^i\chi^j\epsilon^{(a)}\partial_i\partial_jF^{(a)}(\phi)& =0\\
M\partial_i\epsilon^{(a)}F^iF^{(a)}(\phi)& =0\\
M\chi^i\chi^j\partial_i\epsilon^{(a)}\partial_jF^{(a)}(\phi)& =0\\
M\chi^i\chi^j\partial_i\partial_j\epsilon^{(a)}F^{(a)}(\phi)& =0
\end{split}
\end{equation}
which can be solved if and only if
\begin{equation}
MF^{(a)}(\phi)=0.
\end{equation}
In all our examples in this class of spacetime geometries,
we have $M\sim \frac{1}{r}$, where $r$ is some constant characteristic
radius of spacetime. Therefore the constraint is really
\begin{equation}
F^{(a)}(\phi)=0.
\end{equation}
Note that this is well defined globally,
since the combination~(\ref{kwcombo}) is well defined
globally.
Also note that these constraints reduce to the flat
Minkowski spacetime case in the
limit $r\rightarrow\infty$, in which we have no constraint on $F^{(a)}$
and the superpotential is gauge invariant.

We should point out that we are implicitly giving up gauge-invariance of $W$,
since it is not physically meaningful.  Only the linear
combination~(\ref{kwcombo}), namely,
\begin{displaymath}
K \: - \: \frac{3}{2M}W - \: \frac{3}{2\bar{M}}\bar{W},
\end{displaymath}
is physically meaningful.  If we were to
separately demand that $W$ be gauge-invariant, then the resulting
constraint we would obtain would only make sense for those special
K\"ahler transformations that leave $W$ invariant -- which is to say,
none of them. More explicitly, if we were to treat $K$ and $W$ separately
to analyze their individual gauge transformations using the first line
of~(\ref{superspace-gauge-lagrangian}), then the gauge
invariance of~(\ref{superspace-gauge-lagrangian}) leads us to
\begin{equation}  \label{gauge-trans-superpotential}
\begin{split}
\delta W(\phi)&=-M\epsilon^{(a)}F'^{(a)},\\
MF'^{(a)}&=0,
\end{split}
\end{equation}
where $F'^{(a)}=X^{(a)}K+iD^{(a)}$ which is not invariant under
K\"{a}hler transformations. Now the superpotential is gauge invariant,
and we have the constraint $F'^{(a)}=0$. However, in this case to make
sense of the constraint $F'^{(a)}=0$ globally, we need to use the
globally defined K\"{a}hler potential (whose existence is guaranteed by
the trivial K\"{a}hler class on the target space) and demand
that no K\"{a}hler transformation is allowed, which is exactly what we
have been seeing. Therefore, physically we should work with the
combination~(\ref{kwcombo}).  We should note that in this case, in the limit
$r \rightarrow\infty$ which leads to the flat Minkowski spacetime,
we recover the familiar gauge invariance of the superpotential as expected,
since there will be no appearance of the combination~(\ref{kwcombo})
in the Lagrangian~(\ref{superspace-gauge-lagrangian}),
which is reduced to the flat Minkowski spacetime Lagrangian.

Mathematically, there is another way of understanding the constraint
$F^{(a)}=0$.
Recall that the superpotential is a section of an affine bundle
$(\mathcal{O},A)$ over the target space $X$,
therefore we must lift the action of the gauge group to an action
on this affine bundle.  Comparing gauge transformations
of the
superpotential~(\ref{gauge-trans-superpotential}) with
equation~(\ref{equiv-affine}) in the Appendix,
we see that we can describe the infinitesimal group action as
an infinitesimal lift to the affine bundle, described by
\begin{equation}
\begin{split}
&\lambda=1, \\
&\mu=-M\epsilon^{(a)} F'^{(a)}.
\end{split}
\end{equation}
Thus the lifting property~(\ref{equiv-affine-2}) requires for
example $2F'^{(a)} = F'^{(a)}$ or simply $F'^{(a)}=0$,
{\it i.e.} the fact that the superpotential is a section of the affine bundle
$(\mathcal{O},A)$ puts exactly the same constraint
on the geometry of $X$ as derived from gauge invariance.

Now let us discuss the implications of the constraint
\begin{equation}
F^{(a)}\: = \: X^{(a)}\left(K\: - \: \frac{3}{2M}W
\: - \: \frac{3}{2 \bar{M}} \bar{W}\right)\: + \: iD^{(a)} \: = \: 0
\end{equation}
in detail.
Let us begin with the definition of $D^{(a)}$, namely
\begin{eqnarray*}
\partial_i D^{(a)} & = & -i X^{(a) \overline{\jmath}}
\partial_{\overline{\jmath}} \partial_i K,\\
\partial_{\overline{\jmath}} D^{(a)} & = & i X^{(a) i} \partial_i
\partial_{\overline{\jmath}} K.
\end{eqnarray*}
Integrating the equations above, we find that the most general
solution for $D^{(a)}$ is given by
\begin{displaymath}
D^{(a)} \: = \: -i X^{(a) \overline{\jmath}} \partial_{\overline{\jmath}} K'
\: + \: C,
\end{displaymath}
where $K'$ is any Kahler potential ({\it i.e.} $K' = K + f + \overline{f}$
for any holomorphic function $f$), and $C$ is a constant.
Thus the constraint that $F^{(a)} = 0$ is fixing $C = 0$ (and also
partially fixing $K'$).
Physically, this is setting the Fayet-Iliopoulos parameter to zero.

Let us outline some examples, to understand the implication of this.\\
\\
\textbf{Example:} Let the target space $X$ be $\mathbb{C}^n$,
with the standard K\"{a}hler potential $K=\sum_i|z_i|^2$,
and consider an isometry group $U(1)^k$, in which each $U(1)$ acts
by phases on the $z_i$ as
\begin{displaymath}
z_i \: \mapsto \: \lambda^{Q_i^a} z_i.
\end{displaymath}
Then the holomorphic Killing vectors are given by
\begin{equation}
X^{i(a)}=i Q_i^a z_{i}, \: \: \:
X^{\bar{i}(a)}=-i Q_i^a \bar{z}_{\bar{i}}.
\end{equation}
Then the constraining equations tell us
\begin{equation}
D^{(a)}=-\sum_i Q_i^a |z_{i}|^2.
\end{equation}
For example, if there is only one $U(1)$ and all the $Q_i = 1$,
then we are describing a projective space of zero radius.
If there is only one $U(1)$, $n=4$, two charges are $+1$ and two charges
are $-1$, then we are describing a conifold with zero-size small resolution.
\footnote{Note such zero-size effects imply strong coupling in the
nonlinear sigma model.  In this paper we only consider classical
actions, not quantum physics.}

In the examples above, we saw that the quotient had K\"ahler form
of trivial cohomology class, as expected -- after all, the ungauged
theory is only defined on spaces with trivial K\"ahler class, so one
expects the moduli spaces of the gauge theories to have the same
property.

More generally, it is straightforward to check that the constraint
$F^{(a)} = 0$ ensures that the cohomology class of the K\"ahler form
on the quotient is always trivial.  Briefly, the point is that
$D^{(a)} = 0$ if and only if
\begin{displaymath}
X^{(a) \overline{\jmath}} \partial_{\overline{\jmath}} K'
\: = \: 0
\end{displaymath}
which ensures that $K'$ is gauge-invariant\footnote{
Gauge-invariance of a form $\omega$, at least infinitesimally, is the statement
that for a vector field
\begin{displaymath}
X^{(a)} \: = \: X^{(a) i} \partial_i \: +
\: X^{(a) \overline{\imath}} \partial_{\overline{\imath}}
\end{displaymath}
the Lie derivative $L_{X^{(a)}} \omega = 0$.  For the function $K'$,
\begin{displaymath}
L_{X^{(a)}} K' \: = \: X^{(a) i} \partial_i K' \: + \:
X^{(a) \overline{\imath}} \partial_{\overline{\imath}} K'
\end{displaymath}
whose vanishing follows immediately from $F^{(a)} = 0$.
For the K\"ahler form
$\omega$, gauge-invariance $L_{X^{(a)}} \omega = 0$ is easily checked to
be a consequence of the fact that the $X^{(a)}$ are Killing vectors.
} and so descends to the symplectic quotient,
where it becomes a globally defined K\"ahler potential, whose second
derivative is (manifestly) the descent of the restriction of the
K\"ahler form on the original space.

Thus, we see that the constraint $F^{(a)} = 0$ forces the quotient space
to admit a globally-defined K\"ahler potential, as we would naively expect
from properties of ungauged sigma models.

Let us now apply these general argument to some examples from
section~\ref{sect:review}.\\
\\
\\
\textbf{Example 1: AdS$_4$}\\
In components, the superspace Lagrangian~(\ref{superspace-gauge-lagrangian})
gives the Lagrangian on AdS$_4$ spacetime
\begin{equation} \label{gauge-nlsm-AdS4}
\begin{split}
\mathcal{L}=& -g_{i\bar{\jmath}}\mathcal{D}_{\mu}\phi^i\mathcal{D}^{\mu}
\bar{\phi}^{
\bar{\jmath}}-ig_{i\bar{\jmath}}\bar{\chi}^{\bar{\jmath}}\bar{\sigma}^{\mu}
\mathcal{D}_{\mu}\chi^i
-i\bar{\lambda}^{(a)}\bar{\sigma}^{\mu}\mathcal{D}_{\mu}\lambda^{(a)}-
\frac{1}{4}F_{\mu\nu}^{(a)}F^{\mu\nu(a)}-\frac{1}{2}D^{(a)2}\\
& +\sqrt{2}g_{i\bar{\jmath}}(X^{i(a)}\bar{\chi}^{\bar{\jmath}}
\bar{\lambda}^{(a)}+\bar{X}^{\bar{\jmath}(a)}\chi^i\lambda^{(a)})
-\frac{1}{2r}\mathcal{D}_i(K_j+rW_j)\chi^i\chi^j-\frac{1}{2r}
\mathcal{\bar{D}}_{\bar{\imath}}(K_{\bar{\jmath}}+r\bar{W}_{\bar{\jmath}})
\bar{\chi}^{\bar{\imath}}\bar{\chi}^{\bar{\jmath}}\\
& -\frac{1}{r^2}g^{i\bar{\jmath}}(K_j+rW_j)(K_{\bar{\jmath}}+r\bar{W}_{
\bar{\jmath}})+\frac{3}{r^2} (K+rW+r\bar{W})
+\frac{1}{4}\mathcal{R}_{i\bar{\jmath}k\bar{l}}\chi^i\chi^k\bar{\chi}^{
\bar{\jmath}}\bar{\chi}^{\bar{l}},
\end{split}
\end{equation}
where we have set the gauge coupling to one, and
\begin{equation}
\begin{split}
\mathcal{D}_{\mu}\phi^i&=\partial_{\mu}\phi^i-A_{\mu}^{(a)}X^{i(a)},\\
\mathcal{D}_{\mu}\chi^i&=\partial_{\mu}\chi^i-\omega_{\mu}\chi^i+\Gamma^i_{jk}
\mathcal{D}_{\mu}\phi^j\chi^k-A_{\mu}^{(a)}\partial_j X^{i(a)}\chi^j,\\
\mathcal{D}_{\mu}\lambda^{(a)}&=\partial_{\mu}\lambda^{(a)}-
\omega_{\mu}\lambda^{(a)}-
f^{abc}A_{\mu}^{(b)}\lambda^{(c)},
\end{split}
\end{equation}
are the gauge-covariant derivatives,
with $f^{abc}$ being the structure constants of the gauge group $G$,
and $\omega_{\mu}$ being the spin connection on the AdS$_4$ spacetime.
Similar to the ungauged theory, the supersymmetry transformations can be
derived using the gauged supergravity transformations in \cite{wb}
\begin{equation}
\begin{split}
\delta_{\zeta}\phi^i&=\sqrt{2}\zeta\chi^i,\\
\delta_{\zeta}\chi^i&=\sqrt{2}F^i\zeta+i\sqrt{2}
\sigma^{\mu}\bar{\zeta}\mathcal{D}_{\mu}\phi^i,\\
\delta_{\zeta}F^i&=-\frac{\sqrt{2}}{2r}\zeta\chi^i+i\sqrt{2}\bar{\zeta}
\bar{\sigma}^{\mu}\mathcal{D}_{\mu}\chi^i+2iT^{(a)}\phi^i\bar{\zeta}\bar{\lambda}^{(a)},\\
\delta_{\zeta}A_{\mu}^{(a)}&=i(\zeta\sigma_{\mu}\bar{\lambda}^{(a)}+
\bar{\zeta}\bar{\sigma}_{\mu}\lambda^{(a)}),\\
\delta_{\zeta}\lambda^{(a)}&=F_{\mu\nu}^{(a)}\sigma^{\mu\nu}\zeta-
iD^{(a)}\zeta\\
\delta_{\zeta}D^{(a)}&=-\zeta\sigma^{\mu}\mathcal{D}_{\mu}
\bar{\lambda}^{(a)}-\mathcal{D}_{\mu}\lambda^{(a)}\sigma^{\mu}\bar{\zeta}.
\end{split}
\end{equation}

Note that this Lagrangian reduces to the flat Minkowski spacetime case when
$r\rightarrow\infty$ as expected.
Also note that in this Lagrangian
the gaugino is not coupled to
any part of the affine bundle $(\mathcal{O},A)$,
or equivalently, the lifting of the gauge group action to this line bundle
is trivial. This should be compared to the case of $N=1$ supergravity:
there, the gaugino is a section of the Bagger-Witten line bundle,
so that the gauge group action lifts to this line bundle nontrivially,
which leads to the quantization of the
Fayet-Iliopoulos parameter \cite{dist-sharpe,hellerman-sharpe}.\\
\\
\textbf{Example 2: $S^4$}

Let us now Wick rotate to Euclidean space, and consider the case that
the spacetime is $S^4$, as a related example.
Using the values of $M$, $\bar{M}$, $b_{\mu}$ in equation~(\ref{S4}),
the Euclidean Lagrangian is
\begin{equation} \label{gauge-nlsm-S4}
\begin{split}
\mathcal{L}=& g_{i\bar{\jmath}}\mathcal{D}_{\mu}\phi^i\mathcal{D}^{\mu}
\bar{\phi}^{\bar{\jmath}}+ig_{i\bar{\jmath}}\bar{\chi}^{\bar{\jmath}}
\bar{\sigma}^{\mu}\mathcal{D}_{\mu}\chi^i
+i\bar{\lambda}^{(a)}\bar{\sigma}^{\mu}\mathcal{D}_{\mu}\lambda^{(a)}+
\frac{1}{4}F_{\mu\nu}^{(a)}F^{\mu\nu(a)}+\frac{1}{2}D^{(a)2}\\
& -\sqrt{2}g_{i\bar{\jmath}}(X^{i(a)}\bar{\chi}^{\bar{\jmath}}
\bar{\lambda}^{(a)}+\bar{X}^{\bar{j}(a)}\chi^i\lambda^{(a)})
+\frac{i}{2r}\mathcal{D}_i(K_j-irW_j)\chi^i\chi^j+\frac{i}{2r}
\mathcal{\bar{D}}_{\bar{\imath}}(K_{\bar{\jmath}}-ir\bar{W}_{\bar{\jmath}})
\bar{\chi}^{\bar{\imath}}\bar{\chi}^{\bar{\jmath}}\\
& -\frac{1}{r^2}g^{i\bar{\jmath}}(K_j-irW_j)(K_{\bar{\jmath}}-
ir\bar{W}_{\bar{\jmath}})+\frac{3}{r^2} (K-irW-ir\bar{W})
-\frac{1}{4}\mathcal{R}_{i\bar{\jmath}k\bar{l}}\chi^i\chi^k
\bar{\chi}^{\bar{\jmath}}\bar{\chi}^{\bar{l}}.
\end{split}
\end{equation}
The supersymmetry transformations are
\begin{equation}
\begin{split}
\delta_{\zeta}\phi^i&=\sqrt{2}\zeta\chi^i,\\
\delta_{\zeta}\chi^i&=\sqrt{2}F^i\zeta+i\sqrt{2}
\sigma^{\mu}\bar{\zeta}\mathcal{D}_{\mu}\phi^i,\\
\delta_{\zeta}F^i&=-\frac{\sqrt{2}i}{2r}\zeta\chi^i+i\sqrt{2}\bar{\zeta}
\bar{\sigma}^{\mu}\mathcal{D}_{\mu}\chi^i+2iT^{(a)}\phi^i\bar{\zeta}\bar{\lambda}^{(a)},\\
\delta_{\zeta}A_{\mu}^{(a)}&=i(\zeta\sigma_{\mu}\bar{\lambda}^{(a)}+
\bar{\zeta}\bar{\sigma}_{\mu}\lambda^{(a)}),\\
\delta_{\zeta}\lambda^{(a)}&=F_{\mu\nu}^{(a)}\sigma^{\mu\nu}\zeta-
iD^{(a)}\zeta\\
\delta_{\zeta}D^{(a)}&=-\zeta\sigma^{\mu}\mathcal{D}_{\mu}
\bar{\lambda}^{(a)}-\mathcal{D}_{\mu}\lambda^{(a)}\sigma^{\mu}\bar{\zeta}.
\end{split}
\end{equation}
Using our method above, the gauge invariance of the Lagrangian leads to
the following constraining equation
\begin{equation}
F^{(a)}=X^{(a)}(K-irW-ir\bar{W})+iD^{(a)}=0.
\end{equation}
(As in the ungauged theory, since the $W$ terms in
$K-irW-ir\bar{W}$ are not complex conjugates, one might debate whether
the symmetry transformation relating $K$, $W$ should be called a K\"ahler
transformation.)
This constraint effectively makes $D^{(a)}$ complex,
in line with the general observations in \cite{fest-seib} on how
terms breaking superconformal invariance on $S^4$ are complex.
The real part of the constraint implies the
Fayet-Iliopoulos parameter should vanish.

\subsection{$M=\bar{M}=0$, $b_{\mu}\neq 0$}
For the other class of spacetime geometries which are determined by
having nonzero $b_{\mu}$ and $M=\bar{M}=0$, the situation is quite different.
Take the $S^3\times \mathbb{R}$ which is given by~(\ref{S3*R}) as an example:
the superspace Lagrangian in this case is
\begin{equation}  \label{superspace-gauge-lagrangian-S3*R}
\mathcal{L}=\int d^2\Theta  \Big[-\frac{1}{8}(\bar{\mathcal{D}}
\bar{\mathcal{D}}-8R)(K(\Phi^i, \bar{\Phi}^{\bar{\imath}})+
\Gamma(\Phi^i, \bar{\Phi}^{\bar{i}},V))+\frac{1}{16g^2}W^{(a)}W^{(a)}+
W(\Phi^i)\Big]+h.c.,
\end{equation}
where now the ``chiral density'' superfield $\mathcal{E}$
has the property $2\mathcal{E}=1$.
Demanding~(\ref{superspace-gauge-lagrangian-S3*R}) be invariant under
the superspace gauge transformations
\begin{equation} \label{super-gauge-trans-S4}
\begin{split}
\delta K&=\Lambda^{(a)}F^{(a)}+\bar{\Lambda}^{(a)}\bar{F}^{(a)}-
i[\Lambda^{(a)}-\bar{\Lambda}^{(a)}]D^{(a)},\\
\delta \Gamma &=i[\Lambda^{(a)}-\bar{\Lambda}^{(a)}]D^{(a)},\\
\end{split}
\end{equation}
where $F^{(a)}=X^{(a)}K+iD^{(a)}$ (same as in $N=1$ supergravity),
we are led to the result that the superpotential is gauge invariant with
no further constraints on the theory,
just as ordinary supersymmetric gauge theories on Minkowski spacetime.
After eliminating the auxiliary fields, we find the component Lagrangian
\begin{equation} \label{gauge-nlsm-R*S^3}
\begin{split}
\mathcal{L}=& -g_{i\bar{\jmath}}\mathcal{D}_{\mu}\phi^i\mathcal{D}^{\mu}
\bar{\phi}^{\bar{\jmath}}-ig_{i\bar{\jmath}}\bar{\chi}^{\bar{\jmath}}
\bar{\sigma}^{\mu}\mathcal{D}_{\mu}\chi^i
-i\bar{\lambda}^{(a)}\bar{\sigma}^{\mu}\mathcal{D}_{\mu}\lambda^{(a)}-
\frac{1}{4}F_{\mu\nu}^{(a)}F^{\mu\nu(a)}-\frac{1}{2}D^{(a)2}\\
& +\sqrt{2}g_{i\bar{\jmath}}(X^{i(a)}\bar{\chi}^{\bar{\jmath}}
\bar{\lambda}^{(a)}+\bar{X}^{\bar{\jmath}(a)}\chi^i\lambda^{(a)})
-\frac{1}{2}(\mathcal{D}_iW_j)\chi^i\chi^j-\frac{1}{2}(
\mathcal{\bar{D}}_{\bar{\imath}}\bar{W}_{\bar{\jmath}})
\bar{\chi}^{\bar{\imath}}\bar{\chi}^{\bar{\jmath}}\\
& -g^{i\bar{\jmath}}W_j\bar{W}_{\bar{\jmath}}+\frac{1}{4}
\mathcal{R}_{i\bar{\jmath}k\bar{l}}\chi^i\chi^k
\bar{\chi}^{\bar{\jmath}}\bar{\chi}^{\bar{l}}
+\frac{i}{r}(K_i\partial_0\phi^i-K_{\bar{\imath}}\partial_0
\bar{\phi}^{\bar{\imath}})+\frac{1}{2r}g_{i\bar{\jmath}}
\chi^i\sigma_0\bar{\chi}^{\bar{\jmath}},
\end{split}
\end{equation}
which, just as the ungauged case in~(\ref{nlsm-R*S^3}), is almost the same
as the Minkowski spacetime case, with some extra terms coming from the fact
that the spacetime is curved.
Note these extra terms vanish in the limit $r\rightarrow\infty$,
so this theory reduces to the Minkowski case as expected.
(See for example \cite{dsen1,dsen2} for further discussions of rigidly
supersymmetric theories on this spacetime.)

The supersymmetry transformations of this model on $S^3\times \mathbb{R}$ are
\begin{equation}
\begin{split}
\delta_{\zeta}\phi^i&=\sqrt{2}\zeta\chi^i,\\
\delta_{\zeta}\chi^i&=\sqrt{2}F^i\zeta+
i\sqrt{2}\sigma^{\mu}\bar{\zeta}\partial_{\mu}\phi^i,\\
\delta_{\zeta}F^i&=\sqrt{2}\bar{\zeta}^{\dot{\alpha}}
(i\mathcal{D}_{\alpha\dot{\alpha}}\chi^{\alpha}-
\frac{1}{6}b_{\alpha\dot{\alpha}}\chi^{\alpha})\\
\delta_{\zeta}A_{\mu}^{(a)}&=i(\zeta\sigma_{\mu}\bar{\lambda}^{(a)}+
\bar{\zeta}\bar{\sigma}_{\mu}\lambda^{(a)}),\\
\delta_{\zeta}\lambda^{(a)}&=F_{\mu\nu}^{(a)}\sigma^{\mu\nu}\zeta-
iD^{(a)}\zeta,\\
\delta_{\zeta}D^{(a)}&=-\zeta\sigma^{\mu}\mathcal{D}_{\mu}\bar{\lambda}^{(a)}
-\mathcal{D}_{\mu}\lambda^{(a)}\sigma^{\mu}\bar{\zeta}.
\end{split}
\end{equation}

\section{Background principle}

The authors of \cite{allanetal} proposed a ``background principle'':
if a rigid $N=1$ theory on Minkowski spacetime can be quantum-mechanically
coupled to $N=1$ supergravity in a consistent way,
then it should behave
smoothly under deformation from Minkowski spacetime to AdS as classical
theories.  If a theory can be consistently coupled to gravity,
then it should also be possible to consistently formulate it in
a nontrivial background metric.  In particular, cancellation of
(quantum) anomalies in supergravity couplings
is often tied to (classical) consistency
conditions in rigid theories.

In this section we will observe that the same ideas also apply to
gauge theories, and also trivially extend them to all four-manifolds
of the first type discussed in this paper (for which
$M = \overline{M}$ constant,
$b_{\mu} = 0$), not just AdS$_4$.

Let us begin by reviewing the ungauged case, discussed in \cite{allanetal}.
It was observed in \cite{allanetal} that in ungauged theories,
the purely classical constraint on AdS of having a cohomologically trivial
K\"{a}hler form prevents the appearance of gravitional anomaly when one
couples the rigid theory to supergravity.  In detail, start with the
with the six-form anomaly polynomial of $N=1$ supergravity coupled to an
ungauged nonlinear sigma model \cite{dist-sharpe}:
\begin{equation}
\begin{split}
P_{local}=&\phi^*ch_3(X)-\frac{1}{24}p_1(\Sigma)\phi^*c_1(X)\\
&+\phi^*c_1(L)\left(\phi^*ch_2(X)+\frac{21-n}{24}p_1(\Sigma)\right)\\
&+\frac{1}{2}\phi^*\left(c_1(L)^2 c_1(X)\right)+\frac{n+3}{6}\phi^*c_1(L)^3.
\end{split}
\end{equation}
where $\Sigma$ denotes the four-dimensional spacetime,
$L$ denotes the K\"{a}hler (Bagger-Witten)
line bundle over the target space $X$ (the moduli space of the
supergravity),
$\phi: \Sigma \rightarrow X$ denotes the map defining a vev of the bosons
of the theory,
and $n$ is the complex dimension of the target space $X$.
This anomaly polynomial decomposes as a sum
\begin{equation}
P_{local}=P_{global}+\Delta P,
\end{equation}
where
\begin{equation}
P_{global}=\phi^*ch_3(X)-\frac{1}{24}p_1(\Sigma)\phi^*c_1(X),
\end{equation}
is the anomaly polynomial of the rigid nonlinear sigma model, and
\begin{equation}
\begin{split}
\Delta P=&\phi^*c_1(L)\bigg[
\left(\phi^*ch_2(X)+\frac{21-n}{24}p_1(\Sigma)\right)\\
&+\frac{1}{2}\phi^*\big(c_1(L) c_1(X)\big)+\frac{n+3}{6}\phi^*c_1(L)^2\bigg].
\end{split}
\end{equation}
If the K\"{a}hler form is cohomologically trivial, then $c_1(L)=0$.
Thus, if the rigid theory on AdS$_4$
is classically consistent, then coupling to supergravity does
not change the anomaly:  if the rigid theory is anomaly-free,
then so is the theory coupled to supergravity.

In passing, let us make the trivial observation that the computation
above, which \cite{allanetal} originally only applied to AdS$_4$,
also applies to the other four-manifolds of the first type discussed
in this paper (in which $M=\overline{M}$ is constant, $b_{\mu} = 0$).

It remains to ask whether our construction of $N=1$ gauge theory satisfies
this principle. To show this, let's start with the six-form anomaly polynomial
of a $N=1$ supergravity coupled to a gauged nonlinear sigma model, in which
we gauge some global symmetry $G$ of the target space $X$.
We denote this anomaly polynomial as $P_{local}^G$ to distinguish from
the ungauged case above.  From \cite{dist-sharpe},
\begin{equation}
\begin{split}
P_{local}^G=&\phi^*ch_3(T_{vert}\mathcal{M})-
\frac{1}{24}p_1(\Sigma)\phi^*c_1(T_{vert}\mathcal{M})\\
&+\phi^*c_1(\mathcal{L})\left(\phi^*ch_2(T_{vert}\mathcal{M})+\frac{21-n+dim(G)}{24}p_1(\Sigma)\right)\\
&+\frac{1}{2}\phi^*\left(c_1(\mathcal{L})^2 c_1(T_{vert}\mathcal{M})\right)
+\frac{n+3-dim(G)}{6}\phi^*c_1(\mathcal{L})^3.
\end{split}
\end{equation}
In the expression above, $\phi$ is no longer a map $\Sigma \rightarrow X$.
Instead, to define $\phi$, we first pick a principal $G$ bundle over our
four-dimensional spacetime $\Sigma$,
call it $P$.
(The path integral of the gauge theory sums over $P$'s.)
Define
\begin{displaymath}
\mathcal{M} \: \equiv \: (P \times X)/G
\end{displaymath}
which is a bundle over $\Sigma$ with fiber $X$.
Then, $\phi$ is a section of $\mathcal{M}$, {\it i.e.} a map
$\phi: \Sigma \rightarrow \mathcal{M}$ behaving well with respect to the
projection $\mathcal{M} \rightarrow \Sigma$.
Finally, the line bundle $\mathcal{L}$ is defined by taking
the pullback of $L \rightarrow X$ to $P \times X$, and then using the
$G$-equivariant structure to descend to $\mathcal{M} = (P \times X)/G$,
{\it i.e.} schematically, $\mathcal{L} = (\pi_X^*L)/G$.

It is worth emphasizing at this point that the Fayet-Iliopoulos parameters
of the supergravity theory are encoded implicitly in the expression above.
Specifically, they are encoded in $c_1(\mathcal{L})$.
That Chern class manifestly contains information about $c_1(L)$ on $X$,
and in addition, it also contains information about the choice of
$G$-equivariant structure on $L$.  That equivariant structure encodes
the Fayet-Iliopoulos parameters in the supergravity, as described
in \cite{dist-sharpe}.

It will be useful later to understand this in more detail, so let us
consider a simple example.  Suppose that the supergravity moduli space $X$
is a point, so that any Bagger-Witten line bundle $L$ is automatically
trivial, and $c_1(L) = 0$.  The choice of $G$-equivariant structure is then
simply a one-dimensional representation of $G$, {\it i.e.} an action of
$G$ on the one-dimensional fiber $\mathbb{C}$.
In this case, $\mathcal{M} = \Sigma$ and $\mathcal{L}$
is then the line bundle associated to
the principal bundle $P$ via that representation.  If that representation is
nontrivial, then that associated bundle $\mathcal{L} \rightarrow \Sigma$
will vary as $P$ varies, for general $\Sigma$.

In particular, it will be important later to note that the only way to
ensure that $c_1(\mathcal{L}) = 0$ for all choices of $\Sigma$ and $P$'s
is if both $L \rightarrow X$ is trivial, {\it and} the $G$-equivariant
structure on $L$ is also trivial\footnote{
For a nontrivial bundle, there is no meaningful notion of a `trivial'
equivariant structure, but in the special case that the bundle is trivial,
there is a canonical `trivial' equivariant structure.
} (Fayet-Iliopoulos parameters vanish).

In passing, we should note that the fact that in supergravity,
Fayet-Iliopoulos parameters
appear in anomalies, has been discussed elsewhere
in the literature, see for example \cite{fk,efk} for an excellent
description and overview.

Now, let us return to the background principle and the discussion
of anomalies.
Much as in \cite{allanetal}, in the gauged case
the supergravity anomaly decomposes as
\begin{equation}
\begin{split}
&P_{local}^G=P_{global}^G+\Delta P^G\\
&P_{global}^G=\phi^*ch_3(T_{vert}\mathcal{M})-
\frac{1}{24}p_1(\Sigma)\phi^*c_1(T_{vert}\mathcal{M})\\
&\begin{split}
\Delta P^G=&\phi^*c_1(\mathcal{L})\bigg[\left(\phi^*ch_2(T_{vert}\mathcal{M})+\frac{21-n+dim(G)}{24}p_1(\Sigma)\right)\\
&+\frac{1}{2}\phi^*\big(c_1(\mathcal{L}) c_1(T_{vert}\mathcal{M})\big)+\frac{n+3-dim(G)}{6}\phi^*c_1(\mathcal{L})^2\bigg].
\end{split}
\end{split}
\end{equation}
where $P_{global}^G$ is the anomaly of the rigid $G$-gauged
nonlinear sigma model.

As before, $\Delta P^G$ is proportional to $c_1(\mathcal{L})$.
As we noted earlier, to guarantee that $c_1(\mathcal{L})$ vanish,
we must require not only that the $L \rightarrow X$ be
trivial ({\it i.e.} that the target space $X$ has a cohomologically trivial
K\"ahler form), but also that the Fayet-Iliopoulos parameters vanish.

We saw earlier in section~\ref{sect:rigidsusygauge:ads} that
in rigidly supersymmetric gauge theories on four-manifolds of the
first type, including AdS$_4$, we must classically require
both that $X$ have a cohomologically trivial K\"ahler form and that
the Fayet-Iliopoulos parameters vanish.  Thus, these classical
constraints prevent anomalies when coupling to supergravity,
consistent with the background principle.\\
\\
\textbf{Example: $\mathbb{C}P^n$ model}\\
The $\mathbb{C}P^n$ model can be described either by a nonlinear sigma model
with target space $\mathbb{C}P^n$, or by a $U(1)$ gauge theory with $n+1$
chiral superfields of charge $1$. Since $\mathbb{C}P^n$ is compact with a
cohomologically nontrivial K\"{a}hler form, this theory cannot be coupled to
gravity in a consistent way, as discussed above.
Similarly, the four-dimensional gauge theory is anomalous.
Note, on the other hand, in the example from
section~\ref{sect:rigidsusygauge:ads} we showed that our constraint
$F^{(a)}=0$ requires the target space to have zero radius,
{\it i.e.} not really a $\mathbb{C}P^n$ anymore.
Therefore the constraint
$F^{(a)}=0$ is consistent with the background principle, albeit perhaps
trivially so.

\section{Sigma models on stacks and restrictions on nonperturbative sectors}
\label{sect:stacks}

Physically, in supergravity theories we often want to work with
moduli spaces possessing finite group symmetries \cite{seib-may,
hellerman-sharpe,banks-seib}.
These are
described mathematically by stacks \cite{dist-sharpe,hellerman-sharpe},
certain generalizations of
spaces.  (In fact, mathematically most moduli `spaces' are actually
stacks, so to really make contact between the supergravity literature
and string compactifications, one must consider supergravities
containing sigma models on stacks).

Two-dimensional sigma models on stacks have been extensively studied
in the past, and four-dimensional sigma models have been considered
more recently (see \cite{hellerman-sharpe} for a discussion of the
four-dimensional case, and references therein on two dimensions).

As discussed in \cite{hellerman-sharpe} and references therein,
one way to understand sigma models on stacks concretely is to use
the fact that a smooth Deligne-Mumford stack over the complex numbers
can be presented as a quotient $[X/G]$, for $X$ a space and
$G$ a group (which need not be finite, and need not act effectively).
To such a presentation we associate a $G$-gauged sigma model on $X$.

A stack does not uniquely determine such a presentation, but rather
can often be described as $[X/G]$ for several $X$'s and $G$'s.
Thus, one wants to associate stacks to universality classes of
renormalization group flow, rather than to particular quotients
$[X/G]$.  In two dimensions, there are now extensive checks that
renormalization group flow does indeed identify different presentations
of the same stack (see {\it e.g.} \cite{nr,msx,glsm,hhpsa,cdhps}),
though in four dimensions there are suggestions
\cite{hellerman-sharpe} that the same program might not work.
(As an extreme example, since the gauge kinetic term is not fixed,
some references consider Chern-Simons theories to be examples of
sigma models on stacks, see {\it e.g.} \cite{urs}
though the inclination of the authors is to
only consider Yang-Mills-type kinetic terms in such language.)

In any event, as also observed in \cite{hellerman-sharpe}, even if stacks
do not uniquely determine physics, nevertheless they can give some
general insights.  For example, when we describe a stack in terms of
a quotient $[X/G]$, the K\"ahler form on the stack is determined by
both the K\"ahler form on the covering space and also the
Fayet-Iliopoulos parameter.  To get an exact K\"ahler form on the
stack, the naive generalization of the constraints of \cite{allanetal},
requires both an exact K\"ahler form on the covering space $X$ as well
as vanishing Fayet-Iliopoulos parameter -- the two notions are linked.
(Similarly, in the case of $N=1$ supergravity in four dimensions,
integrality of the cohomology class of the K\"ahler form on the stack
requires both integrality of the K\"ahler form on the covering space
$X$ \cite{bagger-witten} as well as an integrality constraint on the
Fayet-Iliopoulos parameter \cite{dist-sharpe}; again, the two
notions are linked.)  In appendix~\ref{app:anoms} we will give another
example, and discuss how
anomalies in supergravities contained gauged nonlinear sigma models
can be described using stacks in a presentation-independent fashion.

In passing, although the Ferrara-Zumino multiplet has not played any
direct role in this paper, this does seem an appropriate spot to mention that
it transforms as a connection on the Bagger-Witten line bundle over the
quotient stack, which summarizes both its transformations across coordinate
patches and its transformation under gauge transformations.
If the Bagger-Witten line bundle on the quotient stack is trivial,
then the Ferrara-Zumino multiplet exists globally.
This requires not only
that the Bagger-Witten line bundle on the covering space be trivial,
but also that the Fayet-Iliopoulos parameter vanish\footnote{
It was observed in \cite{dist-sharpe} that the Fayet-Iliopoulos parameter
in $N=1$ supergravity in four dimensions is really a choice of lift of
$G$ action to the Bagger-Witten line bundle, and as such, it is not always
possible to set it to zero.  However, when the
Bagger-Witten line bundle is itself trivial, there is
a canonical `zero' lift, and the Fayet-Iliopoulos parameter can vanish.
}.

Another perspective on stacks is also sometimes useful.
Stacks can be defined via their incoming maps.
In particular,
sigma models with restrictions on nonperturbative sectors are
precisely examples of sigma models on stacks.  Specifically, the stacks
describing restrictions on nonperturbative sectors are known as gerbes.
Over the last decade,
there has been extensive work on sigma models with restrictions on
nonperturbative sectors (sigma models on gerbes) in both the physics
(see \cite{dist-sharpe,hellerman-sharpe,nr,msx,glsm,hhpsa,cdhps,tonyme,karp1,karp2,me-vienna,me-tex,me-qts})
and especially the math communities (see for example
\cite{cr,agv,cclt,mann,ajt1,ajt2,ajt3,ajt4,t1,gt1,xt1} for a sample).
Such theories have also been discussed implicitly in a number of other
places in the physics literature (see for example
\cite{ed-anton,edgl2,ron-tony,pouliot,poul-strass,strassler,strassler2}).

\section{Conclusions and outlook}
\label{sect:conclusions}

Supersymmetric nonlinear sigma models and gauge theories on curved spacetime,
as we have seen in this paper, have a lot of interesting properties that are
absent in the usual Minkowski spacetime case.
We have discussed some of these properties in this paper, such as the
appearance of the affine bundle structures, the combination of the
K\"{a}hler potential and the superpotential, the constraints that are
imposed by gauge invariance, the vanishing of the Fayet-Iliopoulos
parameter which enforces the noncompactness of the target space,
the general aspects of universality classes of gauged sigma models
as encoded by stacks, and so on.
These theories provide many examples of supersymmetric models on
general spacetime.  There is no doubt that there should be many other
important new phenomenons that yet to be discovered.

One class of possible new phenomenon should arise as instanton effects in
these supersymmetric theories on general spacetime manifolds.
Instantons and their contribution to nonperturbative physics have been studied
extensively in conventional theories in Minkowski spacetime,
therefore one would expect interesting instanton effects could provide many
new nonperturbative effects to quantum field theories on curved spacetime.
We leave the details of these discussion to future work.

\section{Acknowledgements}

We thank A.~Adams, L.~Anguelova,
R.~Bryant, R.~Donagi, H.~Jockers, J.~Lapan, and I.~Melnikov
for useful conversations.
This work was partially supported by NSF grants DMS-0705381, PHY-0755614,
PHY-1068725.

\appendix

\section{Affine bundles and equivariant structures}
\label{app:affine}

In this section we introduce the concept of affine bundle, as well as the
equivariant structure on affine bundles. First we give the definition of
affine space and affine space morphism \cite{affine}\\
\\
\textbf{Definition} Let $V$ be a vector space over some field $k$.
An \emph{affine space} modeled on $V$ is a set $A$,
together with a map $t: V\times A\rightarrow A$ defined by $t(v,a)=v+a$,
which is a free transitive action of V (as an Abelian group under addition)
on $A$.

Intuitively, an affine space is a vector space without origin.
Clearly if we fix an element $a\in A$ to be the origin,
the above definition makes $A$ into a vector space over the field $k$,
which is isomorphic to $V$.\\
\\
\textbf{Example} A vector space $V$ is naturally an affine space modeled
over itself, with the map $t: V\times V\rightarrow V$
given by the natural addition operation.\\
\\
\textbf{Definition} Let $V$ and $V'$ be two vector spaces over the
same field $k$. Let $A$ and $A'$ be affine spaces modeled on $V$ and $V'$
respectively, with corresponding maps $t: V\times A\rightarrow A$
and $t': V'\times A'\rightarrow A'$. An \emph{affine space morphism}
between $A$ and $A'$ is a map $\varphi : A\rightarrow A'$ and a linear
transformation $\tau: V\rightarrow V'$ such that
\begin{equation}
t'(\tau(v),\varphi(a))=\varphi(t(v,a)), \forall a\in A, v\in V.
\end{equation}

Then an affine bundle on a topological space $X$ can be defined as follows.\\
\\
\textbf{Definition} Let $A$ be an affine space modeled on a vector space $V$.
An \emph{affine bundle} $(V,A)$ over $X$ is a fiber bundle
$\pi :(V,A)\rightarrow X$, defined by the following data:
each point $x\in X$ has a neighborhood $U$ and a $U$-isomorphism
$\varphi : U\times A\rightarrow \pi^{-1}(U)$ such that the
restriction $x\times A\rightarrow \pi^{-1}(x)$ is an affine space isomorphism.

In other words, $V$ is an ordinary bundle and $A$ is a $V$-torsor.

Next we consider the equivariant structure on affine bundles.
We restrict our attention to the case of trivial affine line bundle
$\pi: X\times \mathbb{A}^1\rightarrow X$,
which shows up in our discussion of supersymmetric theories on AdS$_4$.
Let $G$ be a group acting on $X$, then the $G$-action on the affine bundle
$\pi: X\times \mathbb{A}^1\rightarrow X$ is given by \cite{affine-lift}
\begin{equation}   \label{equiv-affine}
g(x,a)=(gx, \lambda_g a+\mu_g(x)), \forall g\in G, x\in X, a\in \mathbb{A}^1,
\end{equation}
where $\lambda$ and $\mu$ are functions on $G\times X$ such that for any $g,h\in G$
\begin{equation}  \label{equiv-affine-2}
\begin{split}
& \lambda_{gh}=\lambda_g\cdot \lambda_h, \lambda_e=1,\\
& \mu_{gh}(x)=\lambda_g\cdot\mu_h(x)+\mu_g(hx), \mu_e=0.
\end{split}
\end{equation}
Let $s:X\rightarrow X\times \mathbb{A}^1$ be an $G$-equivariant section of
this affine bundle, defined by $x\rightarrow (x,\sigma(x))$ where
$\sigma\in \mathcal{O}(X)$. Then from the above result we can see
$\sigma$ satisfies \cite{affine-lift}
\begin{equation}
\sigma(gx)=\lambda_g\cdot\sigma(x)+\mu_g(x).
\end{equation}
In this paper we see that the superpotential of supersymmetric theories on
AdS$_4$ is a section of the trivial affine bundle
$\pi: X\times \mathbb{C}^1\rightarrow X$.
Hence if we consider supersymmetric gauge theory on AdS$_4$,
we will have to lift the gauge group action to this affine bundle,
which is given by the gauge transformation of the superpotential.

\section{Anomalies and stacks}
\label{app:anoms}

In the text, we briefly outlined open questions involving
the application of stacks to four-dimensional physics.
Briefly, there is an issue of presentation-dependence, and although
this issue has been resolved in two dimensions, in four dimensional theories
it is a more significant issue.

In this appendix, we will briefly outline how the anomalies in
four-dimensional supergravity theories described in \cite{dist-sharpe}
can be rewritten in a presetation-independent fashion in terms of stacks.
This discussion is somewhat irrelevant to the bulk of the paper,
but is timely and pertinent to the discussion of stacks, so we have placed
it in this appendix.

First, let us recall the setup from \cite{dist-sharpe}.
Instead of a map $\phi: \Sigma \rightarrow X$ from the four-dimensional
spacetime $\Sigma$ to the moduli space $X$ of the supergravity,
we have a section of a bundle $(P \times X)/G$, where $P$ is a
principal $G$ bundle.  $P$ is somewhat arbitrary, in the sense that the
path integral of the gauge theory will sum over choices.

To compare, a map into the quotient stack $\Sigma \rightarrow [X/G]$ is
determined by a principal $G$ bundle $P$ (with connection), together with
a $G$-equivariant map $\tilde{\phi}: {\rm Tot}(P) \rightarrow X$
(by definition of maps into stacks).  Given such $P$ and $\tilde{\phi}$,
define $\phi^{\#}: {\rm Tot}(P) \rightarrow P \times X$ by,
$p \mapsto (p, \tilde{\phi}(p))$.  Note that the $G$ action commutes:
\begin{displaymath}
g \cdot p \: \mapsto \: (g \cdot p, \tilde{\phi}(g \cdot p)) \: = \:
(g \cdot p, g \cdot \tilde{\phi}(p)) \: = \:
g \cdot (p, \tilde{\phi}(p))
\end{displaymath}
hence $\phi^{\#}$ descends to a map ${\rm Tot}(P)/G (=\Sigma)
\rightarrow (P \times X)/G$, which is precisely the map $\phi: \Sigma
\rightarrow \mathcal{M} = (P \times X)/G$ of \cite{dist-sharpe}.

Moreover, it is straightforward to verify that the map $\phi: \Sigma
\rightarrow \mathcal{M}$ which we have just derived, is a section
of $\mathcal{M}$.  Let $\pi: {\mathcal M} \rightarrow \Sigma$ denote
the projection, then note
\begin{displaymath}
\left( \pi \circ \phi \right): \:
p/G \: \mapsto \: (p, \tilde{\phi}(p) )/G \: = \: p/G
\end{displaymath}
hence $\pi \circ \phi$ is the identity.

To make the rest of the identification of the anomaly polynomial,
it is convenient to work with $\tilde{\phi}$  instead of $\phi$.
This gives an anomaly polynomial on ${\rm Tot}(P)$ instead of $\Sigma$.

In this language, $T_{\rm vert} \mathcal{M} = T[X/G]$,
$\mathcal{L}$ lifts to the line bundle $L$ interpreted as a line
bundle on $[X/G]$, and we can use the fact that
\begin{displaymath}
{\rm dim}\, [X/G] \: = \: {\rm dim}\, X \: - \: {\rm dim}\, G \: = \:
n \: - \: {\rm dim}\, G
\end{displaymath}
(as $n \equiv {\rm dim}\, X$).
Then, the anomaly polynomial discussed earlier,
\begin{equation}
\begin{split}
P_{local}^G=&\phi^*ch_3(T_{vert}\mathcal{M})-
\frac{1}{24}p_1(\Sigma)\phi^*c_1(T_{vert}\mathcal{M})\\
&+\phi^*c_1(\mathcal{L})\left(\phi^*ch_2(T_{vert}\mathcal{M})+
\frac{21-n+{\rm dim}(G)}{24}p_1(\Sigma)\right)\\
&+\frac{1}{2}\phi^*\left(c_1(\mathcal{L})^2 c_1(T_{vert}\mathcal{M})\right)
+\frac{n+3-{\rm dim}(G)}{6}\phi^*c_1(\mathcal{L})^3.
\end{split}
\end{equation}
can be rewritten as
\begin{equation}
\begin{split}
P_{local}^G=& \tilde{\phi}^* ch_3(T[X/G]) -
\frac{1}{24}p_1(\Sigma)\tilde{\phi}^*c_1(T[X/G]) \\
&+\tilde{\phi}^*c_1(L)\left(\tilde{\phi}^*ch_2(T[X/G])+
\frac{21-{\rm dim}\,[X/G]}{24}p_1(\Sigma)\right)\\
&+\frac{1}{2}\tilde{\phi}^*\left(c_1(L)^2 c_1(T[X/G])\right)
+\frac{3 + {\rm dim}\,[X/G]}{6}\tilde{\phi}^*c_1(L)^3.
\end{split}
\end{equation}
(Strictly speaking, on stacks one should consider an extension of ordinary
Chern classes, with components on the extra components of the inertia stack.
The application of such components to anomalies is currently under
investigation; here, to be conservative, we only give the part on the
identity component.)
This should be compared to the expression for the anomaly in the
ungauged theory:
\begin{equation}
\begin{split}
P_{local}=&\phi^*ch_3(X)-\frac{1}{24}p_1(\Sigma)\phi^*c_1(X)\\
&+\phi^*c_1(L)\left(\phi^*ch_2(X)+\frac{21-n}{24}p_1(\Sigma)\right)\\
&+\frac{1}{2}\phi^*\left(c_1(L)^2 c_1(X)\right)+\frac{n+3}{6}\phi^*c_1(L)^3.
\end{split}
\end{equation}
Note that if we replaced all occurrences of $X$ in the expression above
with $[X/G]$, then formally we would have recovered the expression for
the anomaly in the gauge theory.  As spaces are special cases of stacks,
this is a good consistency check of the applicability of stacks to
anomalies.

\end{document}